\documentclass[12pt]{article}
\usepackage{epsfig}

\parskip=\medskipamount % space between paragraphs (TeX)
plus.3em minus.5em \abovedisplayshortskip=.5em plus.2em minus.4em
\renewcommand{\thesection}{\arabic{section}}

\catcode`@=11

\@addtoreset{equation}{section} \@addtoreset{equation}{subsection}
\def\theequation{\ifnum\value{section}=0 \arabic{equation}\ignorespaces
\else \ifnum\value{section}=-1 A.\arabic{equation}\ignorespaces
\else \ifnum\value{subsection}=0
\thesection.\arabic{equation}\ignorespaces \else
\thesection.\arabic{subsection}.\arabic{equation}\ignorespaces
                             \fi
                        \fi
                   \fi}

{\catcode`\'=\active \def'{{}^\bgroup\prim@s}}

\catcode`@=12

\newcommand{\bq}{\begin{equation}}
\newcommand{\be}{\begin{equation}}
\newcommand{\fq}{\end{equation}}
\newcommand{\ee}{\end{equation}}
\newcommand{\bqr}{\begin{eqnarray}}
\newcommand{\beqs}{\begin{eqnarray}}
\newcommand{\fqr}{\end{eqnarray}}
\newcommand{\eeqs}{\end{eqnarray}}

\newcommand{\rf}[1]{(\ref{#1})}

\def\bop#1{\setbox0=\hbox{$#1M$}\mkern1.5mu
    \vbox{\hrule height0pt depth.04\ht0
    \hbox{\vrule width.04\ht0 height.9\ht0 \kern.9\ht0
    \vrule width.04\ht0}\hrule height.04\ht0}\mkern1.5mu}
\begin{document}
\thispagestyle{empty}

\begin{flushright}
\begin{tabular}{l}
% TEP- \\
hep-th/0503194 \\
\end{tabular}
\end{flushright}

\vskip .6in
\begin{center}

{\Large\bf  Geometric Solutions to Non-linear Differential
Equations}

\vskip .6in

{\bf Gordon Chalmers}
\\[5mm]
% {\em address \\
%      address \\
% Los Angeles, CA } \\

{e-mail: gordon@quartz.shango.com}

\vskip .5in minus .2in

\end{center}

A general formalism to solve nonlinear differential equations is given.  
Solutions are found and reduced to those of second order nonlinear 
differential equations in one variable.  The approach is uniformized 
in the geometry and solves generic nonlinear systems.  Further properties 
characterized by the topology and geometry of the associated manifolds 
may define global properties of the solutions.  

\setcounter{page}{0}
\setcounter{footnote}{0}
\vfill\break

\section{Introduction}

The systematic solution to non-linear partial differential
equations has prohibited many advances in mathematics and physics.
These equations, contrary to standard theory and linear equations,
appear disparate and unsolvable in the general case.  In the past the 
solutions to these systems of equations have involved many equation  
dependent techniques that are different in various regimes.  A systematic 
formalism to the solutions of these problems is required.  

In this presentation a foundation is given based on geometric methods 
allowing for a uniform treatment of the solutions to these sets of 
non-linear partial differential systems.  The treatment here is to 
give a solution set to these equations and to provide an optimum 
calculus for future improvements.

The general non-linear system is solved for by finding the geodesics 
on an associated manifold.  Even in the generic multi-dimensional 
system, the solution space is found by solving only second order 
non-linear equations in one proper time variable (e.g. the geodesics).  
The reduction in dimensionality is useful for computations and for 
manifesting global properties of the solutions, such as existence and 
behavior in various regimes.  

\section{Setup}

The work begins with the algebraic solution to systems of polynomial
equations as generated in \cite{ChalmersPoly}, involving toric and 
geodesic properties as found for example in \cite{Toric},\cite{EDM2},
\cite{TH}.  The sets of equations are those of the form,

\bqr
 P_c(z_i;x_j) = \sum_{j=1}^m a_{\sigma(j,)}
 \prod_{l=1}^{n_{\sigma(j,l)}} z_{\sigma(j,l)}^{\tilde\sigma{(j,l)}}
  \ ,
\label{polynomial}
\fqr
with $c$ labeling the equation, and $\sigma_c(j,l)$ labeling the
$l$th term of the $c$th equation.  The permutations of the terms
labeled by $\sigma$.  An example set is,

\bqr
  z_1^n + 2 z_2^m z_4^n + 3 z_3^n = 0
\fqr \bqr
  3 z_2^m + 2 z_3^n z_2^m  + z_4^m = 0  \ .
\fqr
The equations in \rf{polynomial} are general and contain, for
example, the well known case pertaining to Fermat's last theorem,

\bqr
P(z_i) = z_1^n + z_2^n - z_3^n
\label{fermatset}
\fqr
Equations with $m=\infty$ are also contained in the set.

The individual terms are accorded to the function(s) solved for
together with the derivatives.   For example, in the Fermat's set
of equations, one would have $z_1=u(w,t)$, $z_2=u_x(w,t)$, and
$z_3=u_t(w,t)$; this pertains to a quantum mechanics problem in
the two-dimensional space of $(w,t)$.

The interest here, however, is in modeling general sets of
(non-linear) differential equations in a uniform manner.
In the interpretation of \rf{polynomial}, the dynamics of the
system is governed by a set of equations with initial conditions
spanning a (potentially singular) manifold.  To set a basic
groundwork, the equations in \rf{polynomial} parameterize a
holomorphically complex manifold as in \cite{ChalmersPoly}.

The spaces considered are (toric) Calabi-Yau, although more
general manifolds are required.  These manifolds have varieties
defined by the equation set in \rf{polynomial}, and may potentially
be singular.  The beginning setup is then, to every non-linear
differential equation, there is an equation describing a fibration
of a Calabi-Yau manifold $M_Z$ over the possibly curved spacetime
$M_X$.  The fibration is general and an example is,

\bqr
P(z;x) = z_1^n + R(x) z_2^m + z_3^n = 0 \ ,
\fqr
with $R(x)$ a generic function describing the $x$ dependence of the
non-linear differential equation.  The other coordinates are $z_1=
u(x,t)$, $z_2=u_x(x,t)$, and $z_3=u_t(x,t)$.  Hence, the fibration
is obvious in this example, as is the differential equation.  Most
general forms are described by,

\bqr
 P_c(z_i;x_j) = \sum_{j=1}^m a_{\sigma(j,)}
 \prod_{l=1}^{n_{\sigma(j,l)}} Q_{(j,l)}(x)
    z_{\sigma(j,l)}^{\tilde\sigma{(j,l)}}
  \ .
 \label{polynomialtwo}
\fqr
These equations are modeled by the flow in a 'product' of $Z\otimes
X$, and are analyzed in this work.

One example in particular is the Navier-Stokes equations,

\bqr
{\partial\over\partial t} u_i + \sum_{i=1}^n u_j {\partial u_i
    \over \partial x_j} = \nu \Delta u_i - {\partial p\over \partial
    x_i} + f_i (x,t) \ ,
\fqr which has a manifold interpretation in terms of
$v_i=z_{(i,0)}$, $v_{(i,j)}={\partial u_i\over \partial x_j}$ with
$j=0,1,\ldots $, and $p(x)$ and $f_i$ general functions in the
space $X$.  The remaining equations is contained in the divergence
condition,

\bqr 
\sum_{i=1}^n {{\partial u_i}\over \partial x_i} = 0 . 
\fqr
These equations are solvable in the approach here.

The fibration $Z\otimes X$ is important to consider first.  First,
the initial conditions are defined by a point in this space.

The solutions are found via flows in the total space $X\otimes Z$.
There are integrability conditions in the direction of the flow
given the derivatives, i.e. $u(x)=Z$ and $u_x(x,t)=\partial_x Z$.
As an example, consider a manifold $M_Z$ spanned by the variables,

\bqr
u, \quad u_x , \quad u_{xx} , \quad u_{xxt} \ ,
\fqr
which are functions of the base space variables $x$ and $t$.  Then
the dynamics is described by the differential equation
$P(u,u_x,u_{xx},u_{xxt})$, with $u(x,t)$ a function of the variables
on the base $R(x,t)$ (e.g. flat space $R^2$).  Label the functions
as,

\bqr
z_1=u , \quad z_2= {\partial u\over x}={\partial z_1\over \partial x},
 \quad z_3={\partial z_2\over\partial x} , \quad
z_4={\partial z_3\over\partial t} \ . \label{intconditions} 
\fqr
The initial conditions $u^o$, $u_x^o$, $u_{xx}^o$, and $u_{xxt}^o$
at $(x^o,t^o)$ generate a point on the fibration $M_Z\otimes X$.
(These initial values must satisfy the differential equations.)

In order to find the values of $u$ (and $u_x$, $u_{xx}$, $u_{xxt}$) at
a different point $(x,t)$, the former variables follow a flow from
the point $(x^o,t^o)$ to $(x,t)$ while preserving the conditions in
\rf{intconditions}.  The path in $M_Z$ generated through the equations
in \rf{intconditions} are integrability conditions on the coordinates
$(x^o,t^o)$ to $(x,t)$.

\section{The case of $P(z_i)=0$.}

There are two cases of interest, when $P(z_i;x_j)=P(z_i)$ and when
there is explicit base dependence.  The former case is analyzed
first. The $z_i$ coordinates are the functions $u$, $u_x$, and
etc. The values from the initial point $z_i^o$ to the final point
$z_i$ are constrained on the manifold $P(z_i)$ and are generated
via a geodesic flow.  Then the coordinates $x_j$ are determined via 
the integrability conditions in \rf{intconditions}; the inclusion of 
an auxiliary complex space $\bar{\cal P}({\bar z}_i)$ is generally required.

The first step, as in \cite{ChalmersPoly}, is to model a space via the 
polynomial equation $P(z_i)=0$ (with possibly the non-holomorphic 
modification of $\bar{\cal P}({\bar z})$.  The geodesics on these space 
will require solutions.

Formally the spaces used in the work are toric Calabi-Yau,
which may have singularities; the general manifolds are holomorphically 
toric, that is, have two polynomials specifying the both the holomorphic 
and nonholomorphic sides.  A modified ${\cal N}=2$ D-term specification 
can be used to find these metrics.  

These polynomials give information about the existence of solutions, 
further from the locus of points in \rf{polynomial}.  In the K\"ahler 
examples, label the space pertinent to
\rf{polynomial} as $M_{P(z_i)}$ and its Riemannian metric as
$g_{\mu\nu}$.  Its K\"ahler so that both
$g_{\mu\nu}=\partial_\mu\partial_\nu \ln \phi(z_i,{\bar z}_i)$ (in
terms of $z$ and ${\bar z}$, $g=g_{i{\bar j}}$) and
$\Gamma_{\rho,\mu\nu}= 1/2 \partial_\rho
\partial_\mu \partial_\nu \ln \phi(z_i,{\bar z}_i)$ hold.  The polynomial sets
of finite degree are modeled by a finite dimensional
Calabi-Yau; the equation sets of infinite degree (i.e. transcendental)
are described by an infinite dimensional manifold.  

The geodesic equation is, with the coordinates $x=(z_i,{\bar z}_i)$,

\bqr
 {d^2 x^\rho \over d\tau^2} + \Gamma^{\rho,\mu\nu}
    {d x_\mu\over d\tau} {d x_\nu\over d\tau} = 0 \ .  
 \label{geodesic}
\fqr
The K\"ahler example admits a complex form with the connection components, 

\bqr 
\Gamma^{\rho,\mu\nu}= 1/2 \partial^\rho
\partial^\mu\partial^\nu \ln \phi(x_\mu,{\bar x}_\nu)  \ .
\label{Chrisstofel} 
\fqr 
The coordinates $x$ contain both the holomorphic and anti-holomorphic 
pieces describing the geometry.  Its complex K\"ahler form is

\bqr
{d\over d\tau} \left\{ {dz^i\over d\tau} +
  {dz_{\bar j} \over d\tau} \partial^{\bar j} \partial^i \ln(\phi)
    \right\}
 - {d^2 z_{\bar j}\over d\tau^2} \partial^{\bar j} \partial^i
 \ln(\phi)
  = 0 \ ,
\label{complexgeodesic}
\fqr
or

\bqr
{d\over d\tau} \left\{ {dz^i\over d\tau} +
  {dz_{\bar j} \over d\tau} g^{{\bar j},i} \right\}
 - {d^2 z_{\bar j}\over d\tau^2} g^{{\bar j},i}
  = 0 \ .
\fqr
These equations are second order in derivatives, but non-linear
because of the K\"ahler potential.  Solving these equations gives
the $z_i$ coordinates as a function of the 'proper time' $\tau$;
the $2n$ initial and final coordinates $z_i^o$ and $z_i^f$
determine the unknowns.

\begin{figure}
\begin{center}
\epsfxsize=12cm
\epsfysize=12cm
\epsfbox{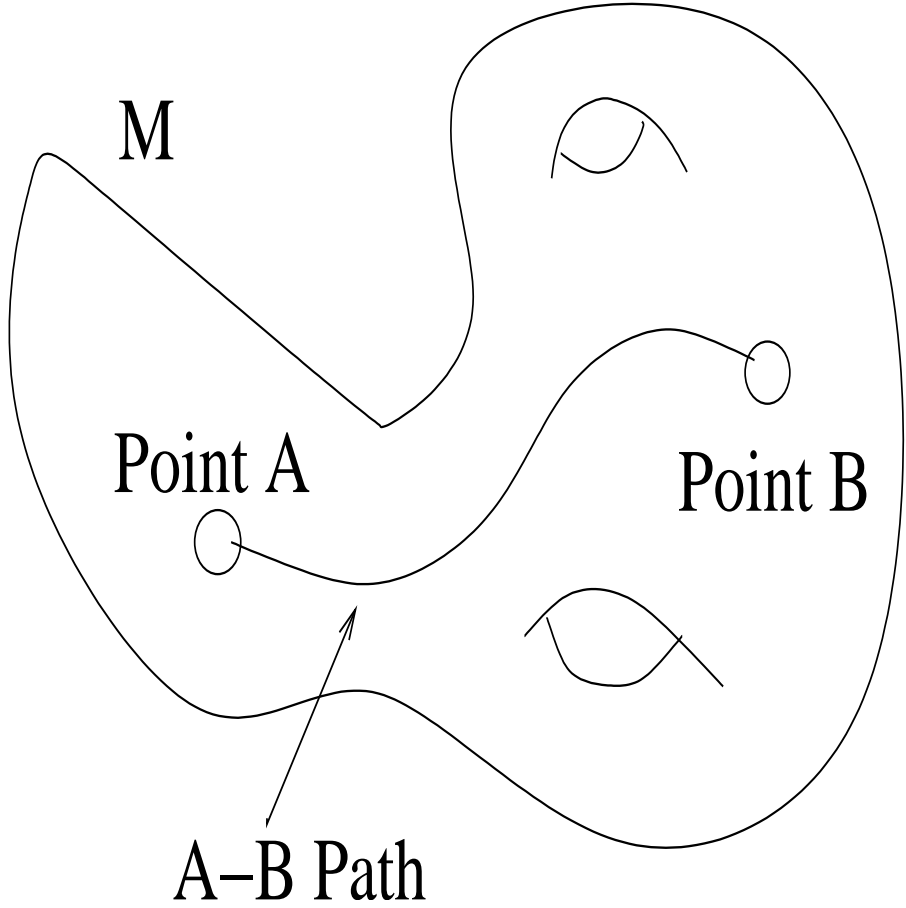}
\end{center}
\caption{The solution via flow in the space $M_Z\otimes M_X$.}
\end{figure}

Once the path from $z_i^o$ to $z_i$ is found, the coordinates in
the base are determined.  As the $z_i$ coordinates are explicit
functions of the initial conditions and $\tau$, the integrability
conditions in \rf{intconditions} are first order in derivatives
and the coordinates $x_j$ requires solutions in $\tau$.  Take
the example of the equation,

\bqr 
u^5 + u_{xx}^4 + u_x = 0 \ . 
\label{samplediffeqn} 
\fqr 
The $z_i$ are $z_1=u(x)$, $z_2=u_{x}(x)$ and $z_3=u_{xx}(x)$.  The
integrability conditions in \rf{intconditions} requires,

\bqr
z_2 = {\partial z_1\over \partial x_j} \ .
\fqr
The function $x$ is found via,

\bqr
{d x\over d\tau} = {1\over z_2(\tau)}
    {\partial z_1 \over \partial\tau} \ .
\label{xsolution}
\fqr
The coordinate in \rf{xsolution} can be solved by an integral

\bqr 
dx = {z_{1,\tau}\over z_2} d\tau 
\fqr 
and gives its $\tau$ dependence.  Further equations need to be solved, 
and these put conditions on the solutions to $z_i$.  For example, the 
equation $z_3=dz_2/dx$ is consistent only if,

\bqr 
{z_{1,\tau}\over z_2} = {z_{2,\tau}\over z_3}
  \qquad \rightarrow \qquad
{1\over 2} {\partial\over\partial\tau} z_2^2 = {\partial z_1\over
 \partial\tau} z_3 \ .
\label{consistence}
\fqr
The equation in \rf{consistence} is satisfied only if the final
coordinates $z_1^f$, $z_2^f$, and $z_3^f$ (and the initial $z_1^o$,
$z_2^o$, and $z_3^o$ at the point $x$) are chosen well.  These
further requirements are {\it algebraic} in nature, and are hand
in hand with the existence of solutions.

\section{Integrability conditions} 

In determining the coordinates $x_j$ from the conditions in 
\rf{intconditions} there are complications due to the number of 
equations and the number of free variables $x_i$.  These complications 
are removed by enlarging the toric space to include a non-holomorphic 
side ${\bar P}({\bar z}_i)=0$, which may be of a large degree.  The 
manifold in principle is no longer K\"ahler, and holomorphically 
separated.  The solution to the non-linear pde's is solved by the 
geodesic flow represented in the holomorphic coordinates $z_i$, while 
the constraints are satisfied via the influence on $z_i$ 
via the flow in the non-holomorpic side ${\bar z}_i$.  This is analyzed 
in this section.  

Consider a differential equation containing,
\bqr
z_1=u \quad z_2=u_{x_1} \quad z_3=u_{x_2} \quad z_4=u_{x_1x_1} \quad
 z_5=u_{x_2x_2} \quad z_6=u_{x_1x_2} \ .
\label{intexample}
\fqr 
There are fewer coordinates $x_i$ than integrability conditions in 
\rf{intexample}.  After solving for the coordinates $x_1(\tau)$ and 
$x_2(\tau)$ with,
\bqr
{dz_1\over dx_1} = z_2 \qquad {dz_1\over dx_2} = z_3 \ ,
\fqr
there are four integrability conditions,
\bqr
{dz_2\over dx_1} = z_4 \quad {dz_2\over dx_2} = z_6
  \quad {dz_3\over dx_1} = z_6 \quad {dz_3\over dx_2} = z_5 \ ,
\fqr
and two relations between them,
\bqr
{dz_2\over z_4} = {dz_3\over z_6} \qquad
{dz_2\over z_6} = {dz_3\over z_5} \ .
\fqr
The integrability conditions, e.g.,
\bqr
{dz_2\over d\tau} ({dx_1\over d\tau})^{-1} = z_4
\fqr
determine four constants, i.e. four of the $z_i^f$ in solving the
geodesic motion.  However, if the solutions are not possible, 
which is very probable, then the final points are taken to be arbitrary 
and a different route (or geodesic) is required; this different route 
changes the functional form of the solution $z_i$ so as to make these 
integrabibility conditions satisfied.  The change in the 
route is made possible by altering the manifold on the non-holomorphic 
side ${\bar P}({\bar z})$, while preserving the holomorphic polynomial 
$P(z_i)$ which models the differential equation.  

In a separate issue, perturbations may be added to make the integrability 
conditions linear in derivatives; this makes the analysis easier in 
solving these equations.  
If the differential equation does not contain the first derivative
in the coordinate, then either: (1) the first non-vanishing term
(e.g. $u_{xx}$ or higher order) must be used to define the spatial
coordinate $x_j$, or (2) a perturbation of the form $u_x$ has to be
added to the differential to define $x_j$ as in the previous.  The
first option requires a second order differential equation to solved,
as opposed to a first order one, $z_2={d^2 z_1/dx^2}$,

\bqr
z_2 = -({\partial x\over\partial\tau})^{-3} {d^2x\over d\tau^2}
 z_{1,\tau} + ({\partial x\over\partial\tau})^{-2} z_{1,\tau\tau} \ .
\fqr
Equations of these types (to be solved for $x$) are more complicated,
and first order linearity is preferred.  The perturbations
$\alpha_i u_{x_i}$ are simpler to add, and post the analysis the
coefficients $\alpha_i$ are taken to zero.  Of course, if the
solutions are singular then the previous method should be used.

The example given in \rf{samplediffeqn} describes how the general
non-linear differential equation is solved for $z_i(\tau;z_i^o,x_j^o)$
and $x_j(\tau;z_i^o,x_j^o)$.  The general solution requires two
non-trivial steps: (1) the solution of a non-linear second-order
differential equation describing the geodesic condition, and (2) the
solution to a set of algebraic conditions describing the integrability
relations.  These steps are uniform to all of the non-linear partial
differential equations that satisfy $P(z_i)=0$, i.e. without any
$x_i$ dependence.

The integrability conditions, if not satisfied via the solution to
the direct equations of motion via the toric variety with complex 
conjugates $P(z_i)=0$, and $P({\bar z}_i)=0$, then a procedure is 
developed based on a different manifold is used.  Consider 
a general space defined by the constraint,

\bqr
 P_c(z_i) = \sum_{\sigma(i)} a_{\sigma(i)} \prod_i z^{\sigma(i)} \ ,
\fqr 
and 
\bqr 
{\bar P}_c({\bar z}_i) = \sum_{{\bar\sigma}(i)}
{\bar a}_{{\bar\sigma}(i)} \prod_i {\bar z}^{{\bar\sigma}(i)} \ ,
\label{nonholomorphic}
\fqr
with possible logarithmic modifications $\ln(z)$ and $\ln({\bar
z})$.  In general, the Laurent expansion with positive and
negative powers of the coordinates span all of the non-analytic
terms.  There are in principle an infinite number of constants in 
the general non-holormpohic side represented in \rf{nonholomorphic}. 
These constants are used to satisfy the integrability conditions.

The point of introducing the coefficients $a_{\sigma}$ and ${\bar
a}_{\sigma}$ is to allow functional dependence in the components
of the Christoffel connection, so as to allow the integrability
conditions (e.g. \rf{intconditions}) to be satisfied.  In general
the metric is a function of these coefficients $a$ and ${\bar a}$.
The holomorphic piece $P(z_i)=0$ is used to describe the analyzed
non-linear differential equation.  The non-holomorphic constraint
${\bar P}({\bar z}_i)$, with the coefficients ${\bar a}$ have
enough terms (an infinite defining a non-analytic function) so as
to allow the constraints to be satisfied.  

As the holomorphic side
of the manifold is parameterized by $P(z_i)$, the final solution
for $Z_f^\mu = Z^\mu(\tau_f)$ will be obtained by a geodesic
condition in the total manifold spanned by $P(z_i)$ and ${\bar
P}({\bar z}_i$, in such a manner that the consistency conditions
are satisfied.  The latter are maintained by choosing the ${\bar
P}({\bar z}_i)$ and the initial and final non-holomorphic
coordinates, ${\bar z}_i$ and ${\bar z}_f$, appropriately to
repeat their solution via the geodesic motion in the total space.  
The solution to the consistency conditions could be complicated 
due to the possibly complicated solutions to the coordinates 
$z_i$, although in only one variable.  
The geodesic equations are non-holomorphic in the sense that both
$z_i^\mu$ and ${\bar z}_i^\mu$ are required (and the mixed indexed
$\Gamma$) in the individual flows; however, the solution to the 
non-linear partial differential equations are obtained from only 
the holomorphic piece.

In the procedure described, the general metric on a non-complex 
(holomorphically toric) manifold is required together with the
solution to the geodesics on the space.  However, the holomorphic
side is described by the original function $P(z_i)$ describing the
non-linear partial differential equation.  The reproduction of the
integrability conditions, given for the one example in
\rf{intconditions}, requires {\it algebraic} solutions to the
parameters ${\bar a}_{\sigma(i)}$ (the solutions may require 
transcendental algebra).

The primary difficulty in the solution is solving for the geodesics
in the Calabi-Yau manifold and also solving the algebraic equations.
However, the treatment of all the non-linearities is reduced to
only second-order partial differential equations in one variable 
(an analysis of algebraic equations is presented in \cite{ChalmersPoly}, 
requiring also second order partial differential equation solutions).

The existence of solutions and the singularities are described by
the steps in the previous example, and are generally given by the
behavior of the metric on the associated manifold.  It is relevant
that the non-linearities and the types of solutions may be described
by the geodesics and properties of the underlying geometry.  The
consistency conditions describe a submanifold connected to the
initial conditions $z_i^o$ and $x_j^o$.

The solutions via this method generate the $\tau$ dependence in
the coordinates $x_j$ and the functions $z_i$, i.e.
$x_j(\tau;x_i^o,z_k^o)$ and $z_i(\tau;x_j^o,z_k^o)$.  The $\tau$
dependence may be solved for as a function of $x_j$ and
substituted in the solution to find $u$ ($u_x$, \ldots), or the
$z_i$, in order to manifest the solution's $x_i$ dependence.

The interpretation of the auxiliary non-holomorphic components
$z_i$ may be put in the form of integrability constants, perhaps
with the the non-holomorphic polynomial ${\bar P}({\bar z})$
having a group theory interpration.

Spatial dependence of the initial conditions should be commented
on.  The initial conditions of $u(x,t)$, $u_x(x,t)$, $\ldots$,
correspond to a point in the manifold parameterized by $P_c(z_i)$.
A spatially dependent 'wave-packet' at time $t=0$ as an initial
condition may be constructed by choosing an appropriate point $Q$
in the manifold and finding the values in a localized region of
space ${\vec x} \in M_i$ at time $t=0$ in which $u({\vec x},0)$ is
given, via the transports from the point $Q$ to the 'final' points
in this region.  The solutions at these 'final' points are evolved
into the required regions $t\neq 0$ and the ${\vec x}$.

\section{The case of $P(z_i;x_j)=0$.}

The primary difference in this case is the inclusion of explicit
coordinate dependence in the set of coupled non-linear partial
differential equations.  For example, there might be a general
term $p(x)$ as in the Navier-Stokes equations,

\bqr
{\partial\over\partial t} u_i + \sum_{i=1}^n u_j {\partial
u_i \over \partial x_j} = \nu \Delta u_i - {\partial p(x_i)\over
\partial
    x_i} + f_i (x,t) \ .
\fqr
More general equations could have the $x$ coordinates spanning a
curved spacetime, with the (coupled set of) non-linear equations
taking values on it and there being explicit metric dependence.

The general $x$ dependence in the equations described by
$P_a(z_i;x_j)$ is incorporated by enlarging the flow in the $z_i$
coordinates to include $x_j$.  In this manner, the constraints
$P_a(z_i;x_j)$ are always preserved. The non-holomorphic
polynomial(s) ${\bar P}_{a({\bar z}_i)}$ describe the remaining
side of the manifold and are used in the geodesic flows as in the
previous case of no explicit x-dependence.

\section{Concluding remarks}

The general set of coupled non-linear differential equations is
described and reduced in terms of flows in manifolds, and the
solution of algebraic equations.  The solution to the differential
equations is described by geodesic flow, that is they reduce to
second order non-linear differential equations in one variable
only.  The multi-variable aspect is further transferred to the
required solving of a set of algebraic equations (of potentially
infinite degree) in an auxiliary set of variables ${\bar a}$.

The second order differentiality in one variable bears a
resemblance to a Lagrangian particle description.  A Hamiltonian
first order description would be useful.  

The nonlinear partial differential equations describes a manifold;
due to the type of manifold, the solutions inherit modular
properties and potentially further ones akin to a mirror symmetry
on toric varieties.  Furthermore, the manifolds' topological
properties potentially may be used to classify solutions.

The geodesic flows, and the sub-manifolds they parameterize,
describe chaotic versus periodic solutions and characterize the
solutions according to initial conditions.  For example, a closed 
geodesic obviously generates non-chaotic behavior; these topological 
properties characterize the integrability, and ergodicity, of the 
solutions.  An ideal situation entails cohomological calculations to 
show integrability properties.  

The obvious use of the proper-time solutions to the non-linear partial 
differential systems, and their form, merits further investigation.  
Also, the connection to the geometry of the manifolds perhaps requires  
input from its topological properties for characterizations of the solutions.  
The explicit form of the geodesics, and the metrics required in order to 
find them, are not presented in this work.


\vskip .3in

\begin{thebibliography}{99}

\bibitem{ChalmersPoly}
G. Chalmers, physics/0503175.  

\bibitem{Toric}
W. Fulton, {\it Introduction to Toric Varieties}, Princeton
University Press, Annals of Mathematics v.131, (1993).

\bibitem{EDM2}
{\it Encylopedic Dictionary of Mathematics}, Iwanamic Shoten
Publishes, Tokyo, 3rd Ed., (1985), English Transl. MIT Press
(1993).

\bibitem{TH}
T. Hubsch, {\it Calabi-Yau manifolds, a bestiary for physicists},
World Scientific Publishing Co, Pte. Ltd. (1992).

\end{thebibliography}
\end{document}